\documentclass[prl,superscriptaddress,preprintnumbers,twocolumn,longbibliography]{revtex4-1} 
\usepackage{amsmath}
\usepackage{amssymb}
\usepackage{amsthm}
\usepackage{amsfonts}
\usepackage{algorithmic}
\usepackage{enumerate}
\usepackage{latexsym}
\usepackage{graphicx}
\usepackage{bm}
\usepackage{subfigure}
\usepackage{setspace}
\usepackage[dvips]{color}
\usepackage{hyperref}
\usepackage{bm}
\begin{document}

\title{Orbital Magnetization under an Electric Field and Orbital Magnetoelectric Polarizabilty for a Bilayer Chern System}
\author{Si-Si Wang}
\affiliation{School of Physics and Materials Science, Guangzhou University, 510006 Guangzhou, China}
\affiliation{School of Mathematics and Information Science, Guangzhou University, 510006 Guangzhou, China}

\author{Yi-Ming Dai}
\affiliation{School of Physics and Materials Science, Guangzhou University, 510006 Guangzhou, China}

\author{Hui-Hui Wang}
\affiliation{School of Physics and Materials Science, Guangzhou University, 510006 Guangzhou, China}
\affiliation{Huangpu Research and Graduate School of Guangzhou University, 510700 Guangzhou, China}

\author{Hao-Can Chen}
\affiliation{School of Physics and Materials Science, Guangzhou University, 510006 Guangzhou, China}

\author{Biao Zhang}
\affiliation{School of Physics and Materials Science, Guangzhou University, 510006 Guangzhou, China}
\affiliation{Huangpu Research and Graduate School of Guangzhou University, 510700 Guangzhou, China}

\author{Yan-Yang Zhang}
\email{yanyang@gzhu.edu.cn}
\affiliation{School of Physics and Materials Science, Guangzhou University, 510006 Guangzhou, China}
\affiliation{Huangpu Research and Graduate School of Guangzhou University, 510700 Guangzhou, China}
\affiliation{School of Mathematics and Information Science, Guangzhou University, 510006 Guangzhou, China}

\date{\today}

\begin{abstract}
In the the real space formalism of orbital magnetization (OM) for a Chern insulator without an external electric field,
it is correct to average the local OM either over the bulk region or over the whole sample. However for a layered Chern insulator in an external electric field, which is directly related to the nontrivial nature of orbital magnetoelectric coupling, the role of boundaries remains ambiguous in this formalism. Based on a bilayer model with an adjustable Chern number at half filling, we numerically investigate the OM with the above two different average types under a nonzero perpendicular electric field. The result shows that in this case, the nonzero Chern number gives rise to a gauge shift of the OM with the bulk region average, while this gauge shift is absent for the OM with the whole sample average. This indicates that only the whole sample average is reliable to calculate the OM under a nonzero electric field for Chern insulators. On this basis, the orbital magnetoelectric polarizablity (OMP) and the Chern-Simons orbital magnetoelectric polarizablity (CSOMP) with the whole sample average are studied. We also present the relationship between the OMP (CSOMP) and the response of Berry curvature to the electric field. The stronger the response of Berry curvature to electric field, the stronger is the OMP (CSOMP). Besides clarify the calculation methods, our result also provides an effective method to enhance OMP and CSOMP of materials.
\end{abstract}

\maketitle

\section{I. Introduction}
In recent years, there has been a significant revival of interest on magnetoelectric effects in materials, including antiferromagnetic systems\cite{antiferro1,antiferro2}, topological insulators\cite{TI1,TI2}, multiferroic composites\cite{multiferro1}. The magnetoelectric effect contributed from the electronic orbital angular momentum is called the orbital magnetoelectric effect (OME), for which some interesting phenomena have been revealed, e.g., orbital Hall effect\cite{OAH1,OAH2,OAH3,OAH4}, orbital torque\cite{OT1,OT2}, in twisted bilayer graphene\cite{TBG1,TBG2,TBG3,TBG4}, helical lattices\cite{HL1,HL2}, and skyrmion crystals\cite{Skyrmion1,Skyrmion2}. One of the most quantities of the OME is the linear orbital magnetoelectric polarizability (OMP), which directly reflects the coupling between the orbital magnetization (OM) and the electric field.

In order to investigate OMP, the first step is to establish a thorough understanding of the OM under an external electric field. The modern theory of OM has achieved this for normal insulators as well as $\mathrm{Z}_{2}$ topological insulators (both with zero Chern number), in the formalism of momentum space ($k$ space) \cite{Xiao2005,Xiao2010,Ceresoli2006,Thonhauser2005,Resta2005,Thonhauser2011,Malashevich2010}. In this context, the OM under an electric field can be expressed as three gauge-invariant constituent terms, namely the local circulation (LC) term, the itinerant circulation (IC) term, and the Chern-Simons (CS) term. The last term is isotropic and proportional to the electric field and therefore vanishes without an electric field. An inevitable problem arises when extending the modern theory to Chern insulators. The Chern-Simons term is gauge dependent so its unambiguous definition needs a smooth gauge. While for a Chern insulator, it is simply impossible to construct a smooth gauge in the entire Brillouin zone\cite{Malashevich2010,Liu2015}.

There is another formalism which expresses the OM and Chern number as local quantities in the real space ($r$ space)\cite{Bianco2011,Bianco2013,Bianco2016,Caio2019,Drigo2020,Sykes2021,Ornellas2022,Hannukainen2022,WeiChen2023}. Furthermore, there are also exciting progresses on the experimental imaging of local magnetic properties\cite{NanoscaleImaging,NanoscaleImaging2}. In the thermodynamic limit, an average of the local quantity over an appropriate region should give correct values consistent with the $k$ space formalism. We have verified previously that in the absence of an external electric field, the total OM averaged over the whole sample coincides with that over the bulk region, and also coincides with the $k$ space formalism, regardless of the Chern number\cite{Wang2022}. This implies that both the $k$ space and $r$ space formalisms of OM are applicable to materials without an external electric field.

For the case of finite electric field, it has been verified that the $k$ space formalism applies equally to OM and its magnetoelectric effect for normal insulators\cite{FiniteElectricField}, while it is not applicable any more to Chern insulators because of the gauge discontinuity of the emerging CS term of OM as mentioned above\cite{Malashevich2010}. Besides the $k$ space formalism, similar questions arise for the $r$ space method: can it be applicable to the case of nonzero Chern number and with an electric field?
The $r$ space formalism of CS term is argued to be the more fundamental definition than the $k$ space one\cite{Olsen2017}. This leads us to suspect that the $r$ space formalism of CS term may be able to circumvent the gauge discontinuity in $k$ space. More specifically, does it need to consider the boundary of the sample in order to circumvent the gauge discontinuity and therewith to obtain a well-defined CS term?

OMP is defined as the orbital contribution to the linear response of magnetoelectric coupling, for example, the response of OM to the external electric field. Corresponding with the OM, OMP can also be written as the sum of similar three gauge invariant contributions originated from the OM formalism\cite{Malashevich2010}. Contributions from the LC term and the IC term can be expressed as the standard linear response of the Bloch functions to the external electric field, denoted as the ``Kubo term'', and the contribution from the Chern-Simons term is called the Chern-Simons orbital magnetoelectric polarizablity (CSOMP)\cite{Liu2015,Coh2011,Essin2010,Olsen2017}. CSOMP has attracted attention due to its relation with topological phases\cite{Qi2011,Turner2012}. This leads to another motivation of this manuscript: finding a method to enhance the total OMP especially the CSOMP.

In this paper, we theoretically investigate the $r$ space formalism of OM for the Chern insulators in the presence of finite electric field, by scrutinizing individual contributions to OM from the bulk and boundary regions. Our calculations are based on a bilayer quantum anomalous Hall (QAH) model with an adjustable Chern number, which has better tunability of properties (band structure, Chern number) than a monolayer one. Such a system is also a candidate for investigation of orbital effects in higher dimensions\cite{Varnava2018,Varnava2020}. We confirm that under nonzero electric field, OM from the bulk region presents a gauge shift with nonzero Chern number which comes from the Chern-Simons term. In other words, contribution from the bulk region to the Chern-Simons term is gauge discontinuous and not well-defined. However, if taking the boundary region into account and averaging over the whole sample, the gauge shift phenomenon will disappear. So OM and its constituent Chern-Simons term from the whole sample average is well-defined. With the whole sample average, both OMP and CSOMP display a close relationship with the response of Berry curvature to the electric field. The stronger the latter, the stronger OMP (CSOMP). This provides a method to estimate and enhance the magnitude of OMP (CSOMP) of materials.

This paper is organized as follows. In Sec.II we present the model, and in Sec. III we lay out the $r$ space method of OM and OMP. In Sec. IV OM and OMP are discussed for the Chern insulator under nonzero electric field. Conclusions are drawn in Sec. V.

\section{II. The Model}
\begin{figure}[htbp]
	\includegraphics*[width=0.53\textwidth]{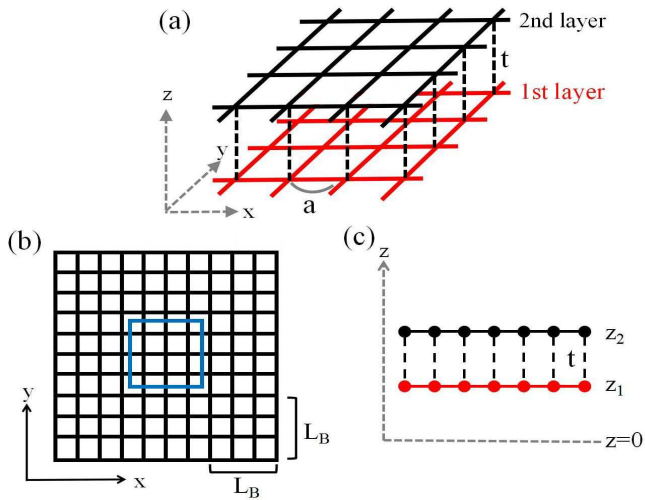}
	\caption{ (a) The lattice structure of our adopted bilayer model. The first layer(red) and the second layer (black) have different model parameters and the same lattice constant $a$. $t$ is the inter-layer coupling strength. (b) The top view of our model. The region enclosed by blue lines represents the bulk region. $L_B$ denotes the thickness of the boundaries. (c) The side view of our model. The $z$ coordinate of the first (second) layer is $z_1$ ($z_2$). }
	\label{Figlattice}
\end{figure}

We choose a bilayer QAH system as the minimum model for our study\cite{Wang2019,Wang2022}, as illustrated in Fig. \ref{Figlattice}. Such topological bilayer systems can be implemented experimentally by using state-of-the-art technologies, such as micro hetero-structures\cite{Fabrication1, Fabrication2, Fabrication3,Fabrication4}, ultracold atomic\cite{FabricationOL1,FabricationOL2,FabricationOL3} or photonic systems\cite{FabricationPhot}.
Our starting point is the spinless Hamiltonian
\begin{equation} \mathcal{H}_{\mathrm{bi}}=\mathcal{H}_1+\mathcal{H}_2+\mathcal{H}_c,\label{EqTotalHamiltonian}
\end{equation}
where $\mathcal{H}_L$ ($L=1,2$) corresponds to the $L$-th layer, and
$\mathcal{H}_c$ is the coupling between them. Each layer is the spin-up component of the Bernevig-Hughes-Zhang model defined on the square lattice, with one $s$ orbital and one $p$ orbital on each site\cite{Bernevig2006}. In the $k$ space, the $L$-th layer Hamiltonian reads
\begin{equation}
	\mathcal{H}_L=\sum_{\bm{k,\alpha\beta}}c^{\dagger}_{L;\bm{k}\alpha}H_{L;\alpha\beta}(\bm{k}) c_{L;\bm{k}\beta},\label{EqHL0}
\end{equation}
where $c^{\dagger}_{L;\bm{k}\alpha}$ ($c_{L;\bm{k}\alpha}$) creates (annihilate) an electron with wavenumber $\bm{k}$ and orbital $\alpha\in \{s,p\}$ in layer $L\in\{1,2\}$. Here, $H_{L;\alpha\beta}(\bm{k})$ is a $2 \times 2$ matrix as\cite{Bernevig2006}
\begin{eqnarray} \label{EqH}
	H_L(\bm{k})&=&\varepsilon_L(\bm{k})I_{2\times 2}+\sum_{i}d^i_L(\bm{k})\sigma_i \label{EqBHZ}\\
	\varepsilon_L(\bm{k})&=&-2D_L\big[2-\cos k_{x}-\cos k_{y}\big] \nonumber \\
	d^1_L(\bm{k})&=&A_L\sin k_{x}, \quad d^2_L(\bm{k})=A_L\sin k_{y}\nonumber\\
	d^3_L(\bm{k})&=&M_L-2B_L\big[2-\cos k_{x} -\cos k_{y}\big],\nonumber
\end{eqnarray}
where $\sigma_i$ are the Pauli matrices acting on the orbital space. The real space version of the $L$-th layer Hamiltonian $\mathcal{H}_L=\sum_{ij,\alpha\beta}c^{\dagger}_{L;i\alpha}H_{L;\alpha\beta}(i,j) c_{L;j\beta}$ can be obtained from Eqs. (\ref{EqHL0}) and (\ref{EqBHZ})
by performing a straightforward inverse Fourier transformation
$c_{L;\bm{k}\beta }=\frac{1}{\sqrt{V}}\sum_{i}c_{L;i\beta }e^{-i\bm{k}\cdot {\bm{r}_{i}}}$, where $i$ is the site index.

In the absence of the inter-layer coupling $H_c$, the band structure and Chern number of each layer can be tuned independently by varying model parameters $A_L$, $B_L$, $D_L$ and $M_L$. For example, the band gap is $2|M_L|$, and the Chern number
\begin{equation}\label{EqCL}
	C_L=
	\left\{
	\begin{array}{lll}
		+1,&\quad 0<M_L/2B_L<2 \\
		0,&\quad M_L/2B_L<0\\
		-1,&\quad M_L/2B_L>2\\
	\end{array}
	\right.
\end{equation}
The inter-layer coupling is illustrated in Fig. \ref{Figlattice} (c). The first layer (red) and the second layer (black) are stacked along the $z$ direction, with the simple
form of inter-layer coupling in real space as
\begin{equation} \label{EqHc}
	\mathcal{H}_c=\sum_{i\alpha}\big(t c^{\dagger}_{1;i\alpha}c_{2;i\alpha}+\mathrm{H. c.}\big)
\end{equation}
where $t$ is the coupling strength.
If the adiabatic turning on of the inter-layer coupling term (\ref{EqHc}) does not close the bulk gap, the Chern number of the bilayer system is just the sum of those of each layer, $C=C_1+C_2$. Otherwise, $C$ will change with varying the band gap $|M_L|$ and the inter-layer coupling strength $t$. Our previous work shows that the Chern number $|C|$ of this bilayer system can take three different value, namely $0$, $1$ and $2$\cite{Wang2019,Wang2022}.

\section{III. Methods}

Applying a static homogeneous electric field $\bm{\varepsilon}$ along $z$ direction (perpendicular to the layer plane), the full Hamiltonian reads
\begin{equation}\label{EqHamiltonianEfield}
	\mathcal{H}=\mathcal{H}^0+e\bm{\varepsilon}\cdot\bm{z}
\end{equation}
where $\mathcal{H}^0$ represents the Hamiltonian in Eq. (\ref{EqTotalHamiltonian}) with $\bm{\varepsilon}=0$. Here $\bm{z}$ denotes the coordinate of each layer along $z$ direction. As shown in Fig. \ref{Figlattice} (c), the coordinate of the first (second) layer is $z_1$ ($z_2$).

In the presence of an external electric field, the total OM of a Chern insulator in $r$-space can be expressed as \cite{Malashevich2010},
\begin{equation}\label{EqMr-nonzero}
	\begin{aligned}
	  &M=M_{\mathrm{LC}}+M_{\mathrm{IC}}+M_{\mathrm{CS}}+M_{\mathrm{BC}}\\
	  &M_{\mathrm{LC}}=\frac{e}{\hbar cA}\mathrm{ImTr}\left\lbrace PxQH_0QyP\right\rbrace \\
  	  &M_{\mathrm{IC}}=-\frac{e}{\hbar cA}\mathrm{ImTr}\left\lbrace QxPH_0PyQ\right\rbrace \\
  	  &M_{\mathrm{CS}}=-\frac{2e^2}{\hbar cA}\varepsilon \mathrm{ImTr}\left\lbrace PxPyPz\right\rbrace \\
	  &M_{\mathrm{BC}}=-\mu\frac{e}{2\pi\hbar cA} \mathcal{C}\\
	  &\mathcal{C}=4\pi \mathrm{ImTr}\left\lbrace QxPyQ\right\rbrace
    \end{aligned}
\end{equation}
where $e$ is the magnitude of the electronic charge, $c$ is the vacuum speed of light and $A$ is the sample area. $P$ is the projection operator onto the ground-state occupied subspace, $P=\sum_{n}\mid\psi_{n}\left\rangle \right\langle \psi_{n}\mid$, and its orthogonal complement $Q$ satisfies $Q=1-P$. Here, $\mid\psi_{n}\rangle$ are the occupied eigenstates. The dimensionless number $\mathcal{C}$ is the integrated Berry curvature up to the Fermi energy, which is quantized as the Chern number $C$ (the topological invariant) if the Fermi energy $\mu$ is in the bulk gap, but may not be quantized if $\mu$ is not in the gap. For convenience in the following, we will always call $\mathcal{C}$ the Chern number, no matter it is quantized or not.

The first three terms of Eq. (\ref{EqMr-nonzero}) are deduced from the OM for normal insulators, with $M_{\mathrm{LC}}$ and $M_{\mathrm{IC}}$ corresponding to the local circulation (LC), the itinerant circulation (IC) motions respectively \cite{Malashevich2010}. With nonzero Chern number, the term $M_{\mathrm{BC}}$ is added in Eq. (\ref{EqMr-nonzero}) to represent the
topological properties of the material, which is proportional to the integrated Berry curvature (BC) up to the Fermi energy.
This term is a direct correspondence to that in the $k$ space formalism in the case of zero electric field\cite{Bianco2011,Bianco2013,Wang2022}. These three terms depend on the electric field $\bm{\varepsilon}$ only implicitly through $P$ and $Q$. In contrast, the Chern-Simons term $M_{\mathrm{CS}}$ gathers contributions with an \emph{explicit} dependence on $\bm{\varepsilon}$. In the limit of vanishing $\varepsilon$, $M_{\mathrm{CS}}$ will becomes zero, and only the rest three constituent components survive\cite{Bianco2011,Bianco2013}.

In the $r$ space formalism (\ref{EqMr-nonzero}), all terms are expressed as $\frac{1}{A}\mathrm{Tr}\cdots$, i.e., averages of some local quantities over the real space. This leads to a noticeable question: over which region? This is conceptually and technologically nonnegligible especially for Chern insulators, where the edge states may have a nontrivial contribution. In the absence of an electric field, this has been resolved that the average region can be chosen either in the bulk (excluding the boundaries) or over the whole sample even for the Chern insulators\cite{Bianco2011,Bianco2013,Wang2022}. Here, as illustrated in Fig. \ref{Figlattice}, the bulk region refers to the inner part of the sample enclosed by the blue dashed square, while the rest part is the boundary region. The whole sample contains both the bulk region and the boundary region.

However, the presence of an external electric field leads to some nontrivial problems which is closely related to the nontrivial nature of magnetoelectric coupling for topologically nontrivial materials\cite{FiniteElectricField,Liu2015,Olsen2017}. Let us summarize the problems we are facing when calculating OM for a layered lattice from Eq. (\ref{EqMr-nonzero}). Now with a nonzero Chern number, the presence of an external electric field $\bm{\varepsilon}$ leads to some ambiguities. First, the CS term has an explicit dependence on $\bm{\varepsilon}$ and $z$, and other terms have an implicit dependence on them. For layered Chern insulators, it was found that the gauge ambiguity can manifests itself as a $z$ coordinate dependent result in the quasi-2D $k$ space formalism\cite{Olsen2017}. Second, in the spatial tracing process, should the contribution from edges be counted in? Previous works show that it is impossible to calculate $M_{\mathrm{CS}}$ in $k$ space for Chern insulators due to its gauge dependence as mentioned above\cite{Malashevich2010,Liu2015}. Is it possible to circumvent this problem in $r$ space? If possible, a further question arises whether the boundaries of the sample need to be considered to obtain a well-defined and single-valued $M_{\mathrm{CS}}$? Therefore in the following, we try to clarify these issues of the $r$ space formalism OM for layered Chern insulators in the presence of an electric field.

The linear magnetoelectric response is defined as $\alpha_{ij}=(\partial{M_{j}}/\partial{\varepsilon_{i}})_{\bm{B}}=
(\partial{P_{i}}/\partial{B_{j}})_{\bm{\varepsilon}}$, where $\bm{M}$ and $\bm{P}$ are the macroscopic magnetization and polarization, respectively. In this manuscript, we focus on the OM so that $\alpha_{ij}$ is called the orbital magnetoelectric polarizability (OMP)\cite{Coh2011}.
Here, the directions of OM and applied electric field $\bm{\varepsilon}$ are both along $z$ direction. So $\alpha_{ij}$ can be simplified as $\alpha_{zz}$ which is denoted as $\alpha$ in the following. By using Eq. (\ref{EqMr-nonzero} ), correspondingly, $\alpha$ can be decomposed as
\begin{equation}\label{Eq-alphaall}
\begin{aligned}	  \alpha&=\alpha_{\mathrm{LC}}+\alpha_{\mathrm{IC}}+\alpha_{\mathrm{CS}}+\alpha_{\mathrm{BC}}\\	  \alpha_{\mathrm{LC}}&=(\partial{M_{\mathrm{LC}}}/\partial{\varepsilon})_{\bm{B}=0}\\
\alpha_{\mathrm{IC}}&=(\partial{M_{\mathrm{IC}}}/\partial{\varepsilon})_{\bm{B}=0}\\
\alpha_{\mathrm{BC}}&=(\partial{M_{\mathrm{BC}}}/\partial{\varepsilon})_{\bm{B}=0}\\
\alpha_{\mathrm{CS}}&=(\partial{M_{\mathrm{CS}}}/\partial{\varepsilon})_{\bm{B}=0}\\
&=-\frac{2e^2}{\hbar cA} \mathrm{ImTr}\left\lbrace PxPyPz\right\rbrace \\
\end{aligned}
\end{equation}

We have checked that the term $\alpha_{\mathrm{BC}}$ is almost zero under a finite but small electric field we adopted in this manuscript. The reason relies on the term $M_{\mathrm{BC}}$ in the OM.
If we average the local OM over the bulk region, $M_{\mathrm{BC}}$ only depends on the topological properties of the sample which is not destroyed by this small electric field\cite{Wang2022}, so $\alpha_{\mathrm{BC}}$ is nearly zero. Besides, our previous work has confirmed that $M_{\mathrm{BC}}$ remains zero with the whole sample average regardless of the Chern number\cite{Wang2022}. $\alpha_{\mathrm{BC}}$ therewith become zero. In other words, the term $M_{\mathrm{BC}}$ will not affect the behavior of the OMP. Hence, we only need to consider the other three constituent terms of OMP even for Chern insulators,
\begin{equation}\label{Eq-alpha}
  \alpha=\alpha_{\mathrm{LC}}+\alpha_{\mathrm{IC}}+\alpha_{\mathrm{CS}}	
\end{equation}
which is consistent with the expression of OMP for normal insulator\cite{Malashevich2010,Coh2011}.
The term $\alpha_{\mathrm{CS}}$ is the Chern-Simons orbital magnetoeletric polarizability (CSOMP)\cite{Coh2011}. It represents the isotropic orbital magnetoelectric coupling, which is equivalent to a term proportional to $\theta_{\mathrm{CS}}\bm{\varepsilon}\cdot\bm{B}$. The dimensionless scalar parameter $\theta_{\mathrm{CS}}$ denotes Chern-Simons axion coupling coupling strength\cite{Wilczek1987,Olsen2017}. The relation between $\alpha_{\mathrm{CS}}$ and $\theta_{\mathrm{CS}}$ satisfies
\begin{equation}\label{theta_CS}
	\alpha_{\mathrm{CS}}=\frac{e^2}{2\pi h}\theta_{\mathrm{CS}}
\end{equation}

In the following, we focus on the behavior of the OM under a finite electric field, in the $r$ space formalism with two different average types. Then the properties of OMP and CSOMP for different topological phases will be investigated.

\section{IV. Results and Discussion}
\subsection{A. Orbital magnetization under a finite electric field}
To investigate OMP, it is necessary to understand the influence of electric field $\varepsilon$ on OM. Previous studies have shown that for the calculation of OM without the electric field, the average of $r$ space formalism over the whole sample is consistent with it over the bulk region even for Chern insulators\cite{Bianco2013,Wang2022}. Does this still work for a nonzero electric field?

\begin{figure}[htbp]
	\includegraphics*[width=0.35\textwidth]{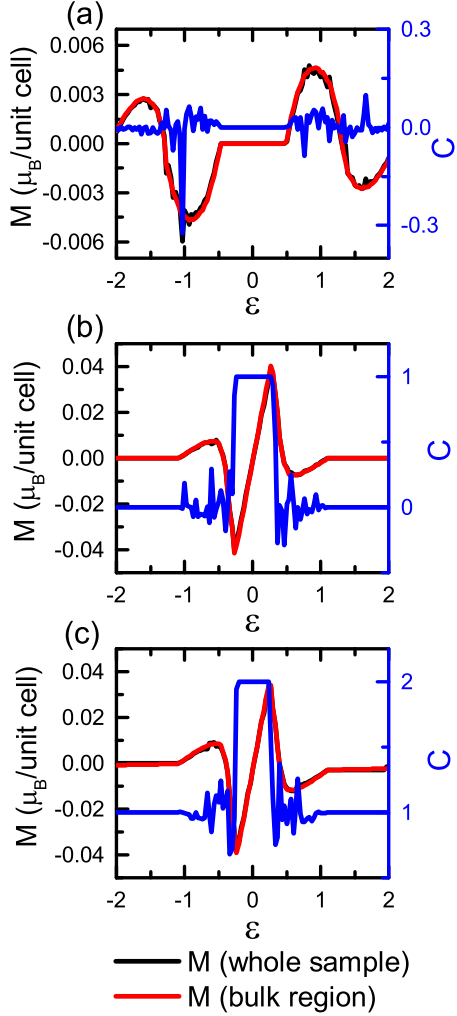}
	\caption{ The OM $M$ with the two different average as a function of the electric field $\varepsilon$ for the bilayer sample with different Chern number. (a)$M_1=M_2=-0.5$, and the bilayer sample with $C=0$. (b) $M_1=-0.5$, $M_2=0.5$, and the bilayer sample with $C=1$. (c) $M_1=M_2=0.5$, and the bilayer sample with $C=2$. Black (Red) lines denotes the $M$ from the whole sample average (the bulk region average). Blue line denotes the the Chern number $C$ of the sample. The Fermi energy $\mu=0.01$. The $z$ coordinate of the first (second) layer is $z_1=1$ ($z_2=2$). The other model parameters: $A_1=A_2=0.3$, $B_1=B_2=0.2$, $D_1=D_2=0$, $t=0.1$. The sample size $N=50$, and the boundry thickness $L_B=\frac{2}{5}N$. }
	\label{FigM-EfieldZ1Z2}
\end{figure}

We first investigate the OM $M$ as a function of the electric field $\varepsilon$, from these two different averages. As shown in Fig. \ref{FigM-EfieldZ1Z2}, black (red) lines correspond to the result from the whole sample (bulk region) average, where $\mu=0.01$ is set in the bulk gap, and the $z$-coordinate of two layers are set as $z_1=1$ and $z_2=2$ respectively. The Chern number (blue line referred to right axis) is also plotted for comparison. Three panels corresponds to samples with different Chern number $C$ at $\varepsilon=0$.
The first observation is that there is a finite window of linear dependence $M(\varepsilon)$ around $\varepsilon$, whose slope is proportional to the Chern number\cite{Ceresoli2006,Olsen2017}.
This linear relationship only holds when the fermi energy is in the bulk gap. When the electric field is strong enough to deform the band structure so that the fermi energy falls into the band. This can be confirmed by the fact that this happens when the Chern number $C$ (blue line and right axis) deviates from integer.

The second feature in Fig. \ref{FigM-EfieldZ1Z2} is, the $M(\varepsilon)$ curve from the whole sample average (black) is completely consistent with (in fact covered by) that from the bulk region average (red), in all three panels. It seems to imply that these two different types of averages are consistent with each other under nonzero electric field even for the Chern insulator. In fact however, it is not always the case. To further clarify this, we study the behavior of OM under different topological phases in more details.

\begin{figure}[htbp]
	\includegraphics*[width=0.46\textwidth]{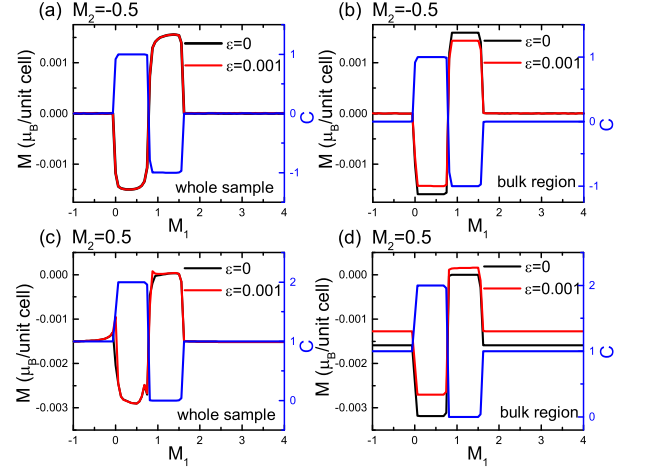}
	\caption{ The OM $M$ from two different average as a function of  $M_1$. Left column: the OM $M$ from the whole sample average. Right column: the OM $M$ from the bulk region average. First row: $M_2=-0.5$ so that $C=C_1+0$. Second row: $M_2=0.5$ so that $C=C_1+1$. Black line: the OM $M$ under zero electric field. Red line: the OM $M$ under nonzero electric field. Blue line: the Chern number $C$ as a function of  $M_1$. The other model parameters are the same as Fig. \ref{FigM-EfieldZ1Z2}. }
	\label{FigM-M1-Z1Z2}
\end{figure}

As mentioned above, if the inter-layer coupling strength $t$ is not strong enough to close the bulk gap, $C$ is just the sum of those of each layer, $C=C_1+C_2$. By varying model parameters, e.g. $M_{L}$, there will be various combinations of $C_L$ ($L\in(1,2)$) through Eq. (\ref{EqCL}). Now we investigate the development of the OM under such changes. Let us first look at the OM $M$ as a function of $M_1$ which determines $C_1$, as displayed in Fig. \ref{FigM-M1-Z1Z2}. Here the first (second) row corresponds to the case of the Chern number $C=C_1+0$ ($C=C_1+1$), and the left (right) column corresponds to the OM from whole sample (bulk region) average, with the black (red) line for the case of an electric field $\varepsilon=0$ ($\varepsilon=0.001$).
The behavior of OM $M$ has a strong correlation with that of the total Chern number (blue line with reference to the right axis) $C$ up to an opposite sign.  Meanwhile, a large OM $M$ occurs only when the Chern number $C$ is nonzero.

Since we have chosen a small electric field strength $\varepsilon=0.001$, the change of $M$ under nonzero electric field $M(\varepsilon=0.001)-M(\varepsilon=0)$ is supposed to be very small compared to $M$ itself. This is the case by averaging over the whole sample as shown in the left column of Fig. \ref{FigM-M1-Z1Z2}, where black and red lines are indistinguishable.
However, this does not apply to the bulk region average (right column). For example, Fig. \ref{FigM-M1-Z1Z2} (b) shows there is a shift of OM $M$ from the bulk region average (red line) under $\varepsilon=0.001$ when $C$ (blue line) is nonzero. We define
\begin{equation}
\triangle{M}_{\mathrm{s}}\equiv {M(\varepsilon)}-{M(0)} \label{EqDeltaMs}
\end{equation}
as a measure of this variation caused by the electric field. As shown in Fig. \ref{FigM-M1-Z1Z2} (b), it satisfies $\triangle{M}_{\mathrm{s}}\propto{\mathrm{sgn}(C)}\frac{\varepsilon}{2\pi}$ with $C=C_1+0$ \cite{Ceresoli2006,Olsen2017}. If we further change the Chern number to $C=C_1+1$, as shown in Fig. \ref{FigM-M1-Z1Z2} (d), $\triangle{M}_{\mathrm{s}}$ becomes larger with $C=2$ than with $C=1$. But we have checked that the former is not twice as large as the latter. Moreover in Fig. \ref{FigM-M1-Z1Z2} (d), this difference $\triangle{M}_{\mathrm{s}}$ is not zero even when $C$ is zero. These indicate that the relationship between $\triangle{M}_{\mathrm{s}}$ and $C$ is not a simple linear relationship with the condition of nonzero $\varepsilon$.
We numerically compare the value of $\triangle{M}_{\mathrm{s}}$ under different Chern numbers. For example, $\triangle{M}_{\mathrm{s}}=\frac{3\varepsilon}{2\pi}$ with $C=2$, and $\triangle{M}_{\mathrm{s}}=\mathrm{sgn}(C)\frac{\varepsilon}{2\pi}$ with $C=\pm1$. There are two different results of $\triangle{M}_{\mathrm{s}}$ for $C=0$. With the condition of $C=0+0$, $\triangle{M}_{\mathrm{s}}$ becomes $0$ in Fig. \ref{FigM-M1-Z1Z2} (b). But if $C=0$ is caused by $C=-1+1$, $\triangle{M}_{\mathrm{s}}=\frac{\varepsilon}{2\pi}$ with $M_1\in(0.8,1.5)$ in Fig. \ref{FigM-M1-Z1Z2} (d). Hence, we conjecture that $\triangle{M}_{\mathrm{s}}$ is not only related to $C$, but also to the specific Chern number of each layer $C_L$.

\begin{figure}[htbp]
	\includegraphics*[width=0.48\textwidth]{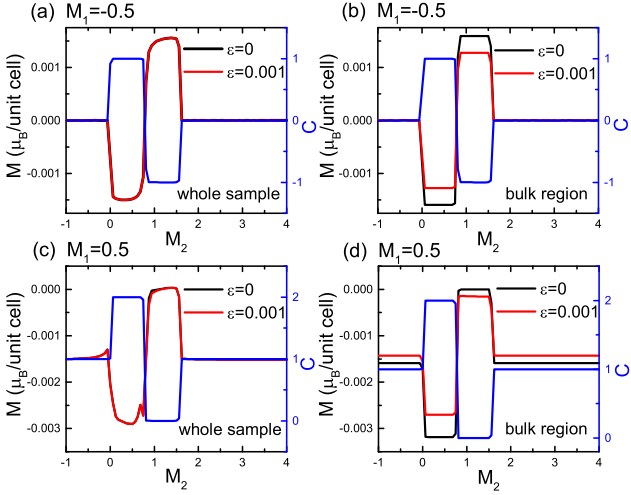}
	\caption{ Similar to Fig. \ref{FigM-M1-Z1Z2} but for the bilayer as a function of  $M_2$. (a) and (b) $M_1=-0.5$. (c) and (d) $M_1=0.5$. Black line: the OM $M$ under zero electric field. Red line: the OM $M$ under nonzero electric field. Blue line: the Chern number $C$ as a function of  $M_2$. Other model parameters are the same as Fig. \ref{FigM-EfieldZ1Z2}. }
	\label{FigM-M2-Z1Z2}
\end{figure}

Similar to Fig. \ref{FigM-M1-Z1Z2} associated with varying $C_1$ from $M_1$, we show results associated with varying $C_2$ in Fig. \ref{FigM-M2-Z1Z2}, where the topological edge states are contributed from the second layer. Still, the first (second) row corresponds to the case of the Chern number $C=0+C_2$ ($C=1+C_2$), and the left (right) column corresponds to that the from whole sample (bulk region) average, with the black (red) line for the case of an electric field $\varepsilon=0$ ($\varepsilon=0.001$).
Similar to the above case of changing $M_1$, $M$ from the whole sample average exhibits a very slight difference under a small electric field as shown in Fig. \ref{FigM-M2-Z1Z2} (a) and (c). From Fig. \ref{FigM-M2-Z1Z2} (b), we can see $\triangle{M}_{\mathrm{s}}=\mathrm{sgn}(C)\frac{2\varepsilon}{2\pi}$ with $C=0\pm1$. Besides, when $M_2$ takes a value between $0.8$ and $1.5$ in Fig. \ref{FigM-M2-Z1Z2} (d), $\triangle{M}_{\mathrm{s}}=-\frac{\varepsilon}{2\pi}$ with $C=0$ ($C=1+(-1)$).
\begin{table}
  \includegraphics*[width=0.45\textwidth]{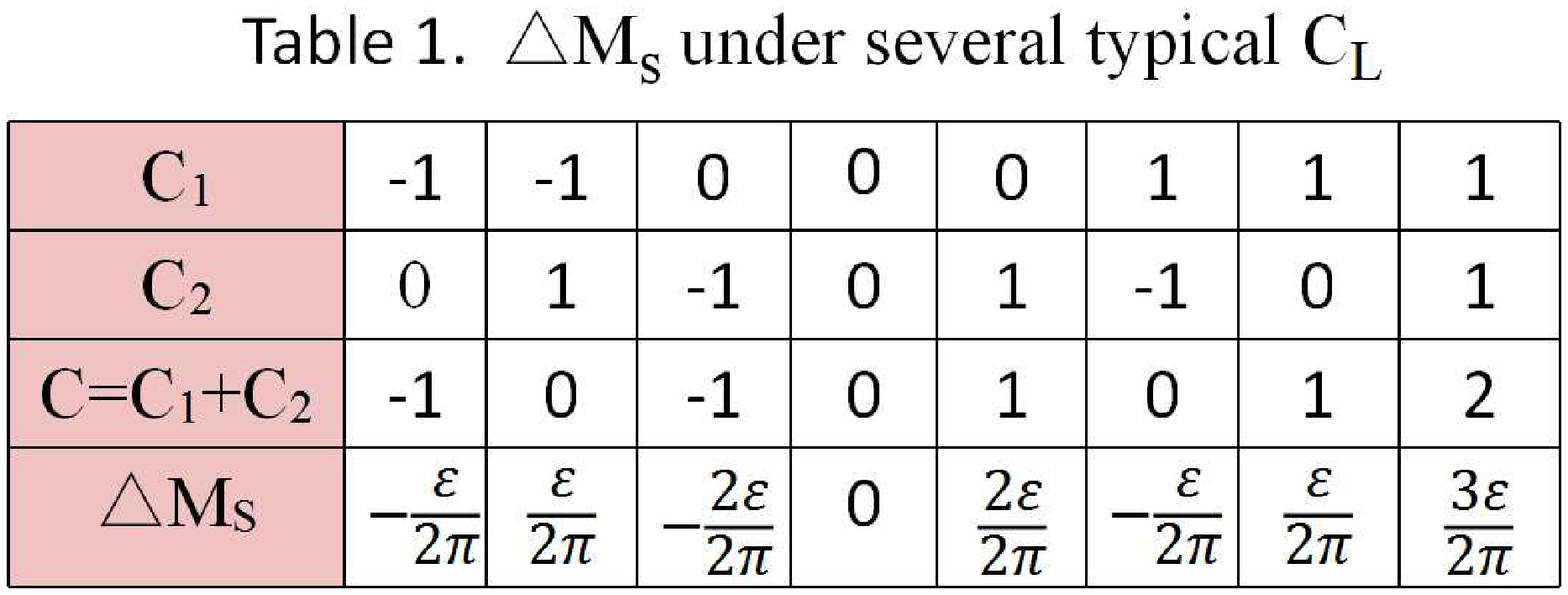}	
	\caption{$\triangle{M}_{\mathrm{s}}$ under different combination of $C_L$.  }
	\label{Tab1}
\end{table}

\begin{figure}[htbp]
	\includegraphics*[width=0.48\textwidth]{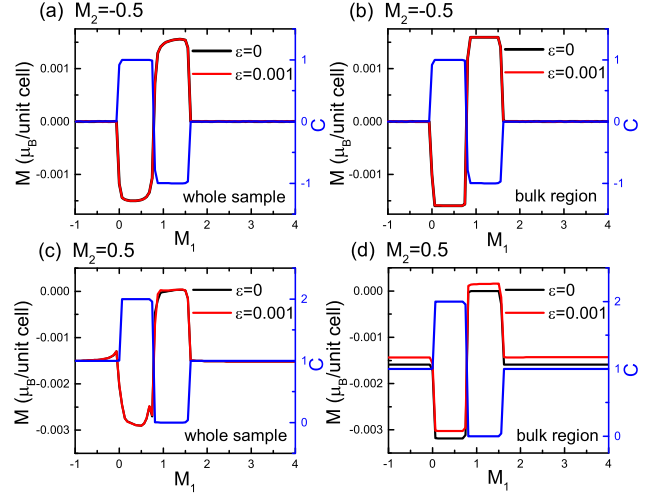}
	\caption{ Similar to Fig. \ref{FigM-M1-Z1Z2} but for the bilayer with coordinate $z_1=0$ and $z_2=1$. (a) and (b) $M_2=-0.5$. (c) and (d) $M_2=0.5$. Black line: the OM $M$ under zero electric field. Red line: the OM $M$ under nonzero electric field. Blue line: the Chern number $C$ as a function of  $M_1$. The other model parameters are the same as Fig. \ref{FigM-M1-Z1Z2}. }
	\label{FigM-M1-Z0Z1}	
\end{figure}

\begin{figure}[htbp]
	\includegraphics*[width=0.48\textwidth]{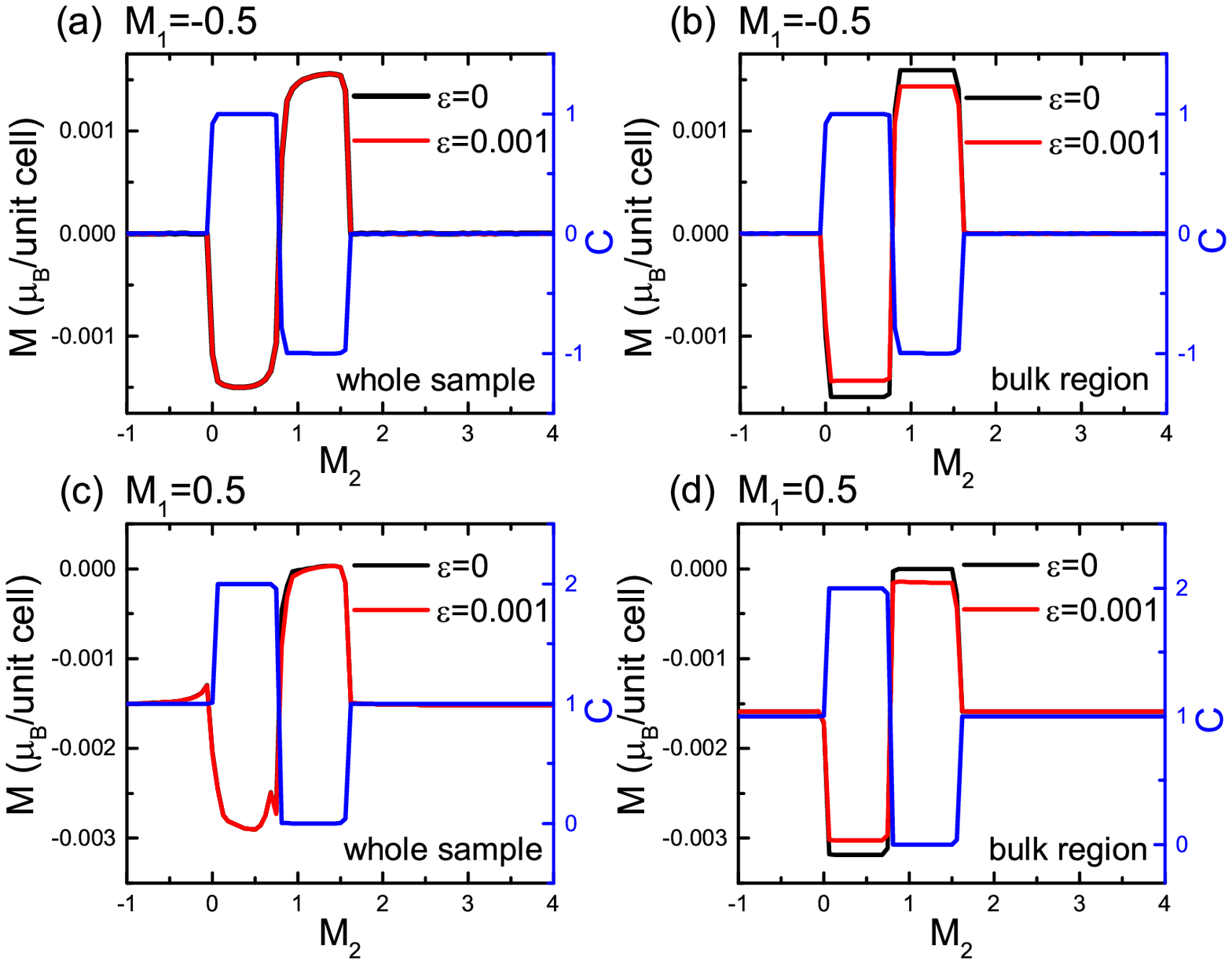}
	\caption{ Similar to Fig. \ref{FigM-M2-Z1Z2} but for the bilayer with coordiante $z_1=0$ and $z_2=1$. (a) and (b) $M_1=-0.5$. (c) and (d) $M_1=0.5$. Black line: the OM $M$ under zero electric field. Red line: the OM $M$ under nonzero electric field. Blue line: the Chern number $C$ as a function of  $M_2$. Other model parameters are the same as Fig. \ref{FigM-M2-Z1Z2}. }
	\label{FigM-M2-Z0Z1}	
\end{figure}

We summarize values of $\triangle{M}_{\mathrm{s}}$ under these typical configurations of $C_L$ in Tab. \ref{Tab1}. It indicates that the contribution of the first layer to $\triangle{M}_{\mathrm{s}}$ is $C_1\frac{\varepsilon}{2\pi}$ while the second layer is $C_2\frac{2\varepsilon}{2\pi}$. Therefore, we conjecture $\triangle{M}_{\mathrm{s}}$ is also related to the $z$ coordinate of each layer, and suppose
\begin{equation}\label{EqdeltaM}
\triangle{M}_{\mathrm{s}}=\frac{\varepsilon}{2\pi}(C_1z_1+C_2z_2)
\end{equation}
where $z_1$ ($z_2$) is the $z$ coordinate of the first (second) layer. It is known that with nonzero Chern number, the gauge ambiguity may lead to a $z$ coordinate dependent result\cite{Liu2015,Olsen2017}. Therefore in order to further confirm this conclusion, we shift the origin of the $z$ coordinate, so that $z_1=0$ and $z_2=1$ for two layers respectively. Corresponding results for the case of varying $C_1$ and $C_2$ are presented in Fig. \ref{FigM-M1-Z0Z1} and Fig. \ref{FigM-M2-Z0Z1}, respectively.


According to our supposed Eq. \ref{EqdeltaM}, with $z_1=0$, the change of $C_1$ will not lead to a nonzero $\triangle{M}_{\mathrm{s}}$, and only $C_2$ can affect
the value of $\triangle{M}_{\mathrm{s}}$. Now this is confirmed from Fig. \ref{FigM-M1-Z0Z1} (b) that $\triangle{M}_{\mathrm{s}}$ is zero as a function of $M_1$. If $C_2=1$ in Fig. \ref{FigM-M1-Z0Z1} (d), $\triangle{M}_{\mathrm{s}}$ remains $\triangle{M}_{\mathrm{s}}=\frac{\varepsilon}{2\pi}C_{2}z_{2}=\frac{\varepsilon}{2\pi}$ regardless of how $M_1$ (i.e. $C_1$) changes. The same is true in Fig. \ref{FigM-M2-Z0Z1} (b) and (d).
Similarly, if the $z$ coordinate is shifted to make $z_1=0$ and $z_2=1$, $\triangle{M}_{\mathrm{s}}$ will disappear with the whole sample average of OM $M$ as illustrated in the left column of Fig. \ref{FigM-M1-Z0Z1} and Fig. \ref{FigM-M2-Z0Z1}.

\begin{figure}[htbp]
	\includegraphics*[width=0.48\textwidth]{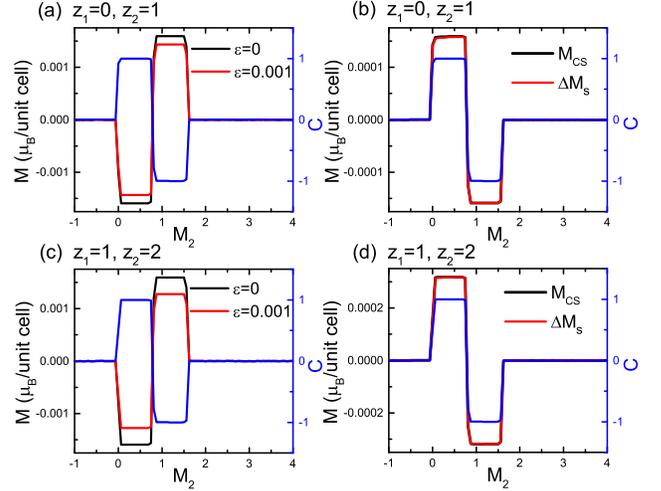}
	\caption{ The OM with the bulk region average as a function of $M_2$. First row: $z$ coordinate for the bilayer as $z_1=0$, $z_2=1$. Second row: $z_1=1$, $z_2=2$. Left column: The OM $M$ from the bulk region average under zero (Black line) and nonzero (Red line) electric field. Right column The constituent component $M_{\mathrm{CS}}$ (Black line) and $\triangle{M}_{\mathrm{s}}$ (Red line) under the electric field $\varepsilon=0.001$. Blue line: the Chern number $C$ as a function of  $M_2$. The other model parameters are the same as Fig. \ref{FigM-M2-Z1Z2}. }
	\label{FigM&Mcs}	
\end{figure}

To figure out the microscopic origin of nonzero $\triangle{M}_{\mathrm{s}}$, we further divide the OM $M$ into four constituent components according to Eq. (\ref{EqMr-nonzero}). It turns out that the $M_{\mathrm{CS}}$ term is the dominating contribution of the nonzero $\triangle{M}_{\mathrm{s}}$, as will be seen in the following.
Let us first focus on the bulk region average as shown in Fig. \ref{FigM&Mcs}. We choose two different types of $z$ coordinate for the bilayer sample: $z_1=0$, $z_2=1$ in the first row of panels, and $z_1=1$, $z_2=2$ in the second row.
Comparing $\triangle{M}_{\mathrm{s}}$ with all constituent terms of OM, we find that $\triangle{M}_{\mathrm{s}}$ as a function of  $M_2$ coincides with the constituent term $M_{\mathrm{CS}}$ in Fig. \ref{FigM&Mcs} (b) and (d). Therefore, $\triangle{M}_{\mathrm{s}}$ comes mostly from the Chern-Simons orbital magnetoelectric coupling $M_{\mathrm{CS}}$. This is consistent with Ref. \cite{Olsen2017}, which shows that due to the gauge discontinuity for Chern insulators, $\theta_{\mathrm{CS}}$ in $k$ space presents a shift under the coordinate shift as $z\rightarrow z+\Delta{z}$, i.e., $\Delta\theta_{\mathrm{CS}}\propto C\Delta{z}$, with $C$ the Chern number. On this basis, we further show the relationship between $M_{\mathrm{CS}}$ and the Chern number $C_L$ of each layer for the bilayer Chern insulator in details. It is reasonable to extend this relationship to the multi-layer Chern insulator as $\triangle{M}_{\mathrm{s}}=\frac{\varepsilon}{2\pi}(C_1z_1+C_2z_2+C_3z_3+...)$.

\begin{figure}[htbp]
	\includegraphics*[width=0.48\textwidth]{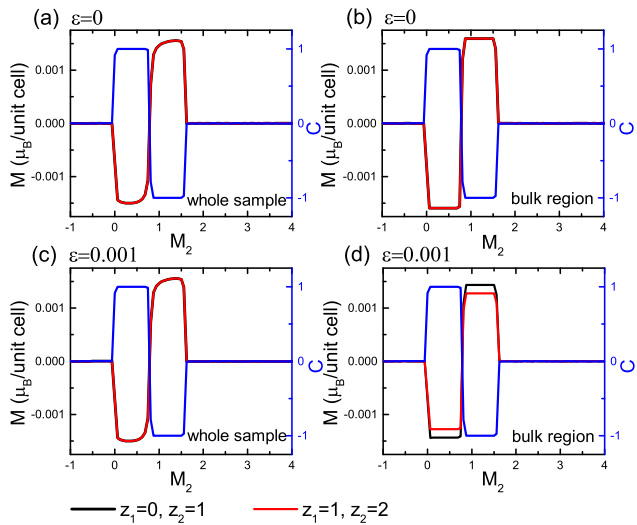}
	\caption{ The OM $M$ as a function of  $M_2$ under zero and nonzero electric field. Left column: The OM $M$ from the whole sample average. Right column: $M$ from the bulk region average. First row: under zero electric field. Second row: under the electric field $\varepsilon=0.001$. Black line: the OM $M$ for the bilayer system with $z_1=0$ and $z_2=1$. Red line: the OM $M$ for the bilayer system with $z_1=1$ and $z_2=2$. Blue line: the Chern number $C$ as a function of  $M_2$. $M_1=-0.5$, and other model parameters are the same as Fig. \ref{FigM-M2-Z1Z2}. }
	\label{FigM-Efield}	
\end{figure}

Since $\triangle{M}_{\mathrm{s}}$ ($\triangle{M}_{\mathrm{s}}\equiv {M(\varepsilon)}-{M(0)}$) from the bulk average is dependent on the $z$ coordinate of the layered system as shown above, it suggests that the OM $M$ from the bulk average is also dependent on the $z$ coordinate of the sample. As presented in Fig. \ref{FigM-Efield}, we summarize the effects from different electric fields, different averaging processes, and different definitions of the $z$ coordinate, with varying $C_2$. The right column denotes the OM $M$ from the bulk average. In Fig. \ref{FigM-Efield} (b), the OM $M(0)$ is not affected by shifting the $z$ coordinate of the sample. However, the OM $M(\varepsilon=0.001)$ shows a shift under two different definitions of the $z$ coordinate as presented in Fig. \ref{FigM-Efield} (d). In other words, the OM $M$ from the bulk region average is gauge invariant only at zero electric field but is not well-defined under nonzero electric field for the Chern insulators.

Now if we turn attention to the whole sample average as illustrated in the left column of
Fig. \ref{FigM-Efield}, the OM $M$ from the whole the whole sample average is immune to the shifting of the $z$ coordinate regardless of whether the electric field $\varepsilon$ is zero or not. It indicates the OM
from the whole the whole sample average is well-defined even for the Chern insulators under a nonzero electric field. This conclusion re-confirms the previous argument that the $r$ space formalism of $M_{\mathrm{CS}}$ is likely to be a more fundamental definition because it is free from gauge variance\cite{Olsen2017}.

In a word, for a layered lattice with nonzero Chern insulator (integrated Berry curvature), the OM $M$ from the bulk region average is not well-defined under a nonzero electric field, due to the gauge discontinuity of the term $M_{\mathrm{CS}}$. Under a nonzero electric field and for a layered lattice with a nonzero Chern number, the correct algorithm to calculate the OM is the whole sample average.

\subsection{B. Orbital magnetoelectric polarizability}

Based on above discussions about the OM of a bilayer, now we can investigate the properties of OMP with different combinations of Chern number $C_L$.
Starting from Eq. (\ref{Eq-alpha}), the total OMP $\alpha$ can also be divided into four constituent terms. The term $\alpha_{\mathrm{BC}}$ under two average processes remains zero as mentioned above, so
only the other three terms have to be considered.
Here, our main focus is the whole OMP $\alpha$ and the constituent CSOMP term $\alpha_{\mathrm{CS}}$ due to its close relation with topological phases\cite{Turner2012}.
\begin{figure}[htbp]
	\includegraphics*[width=0.48\textwidth]{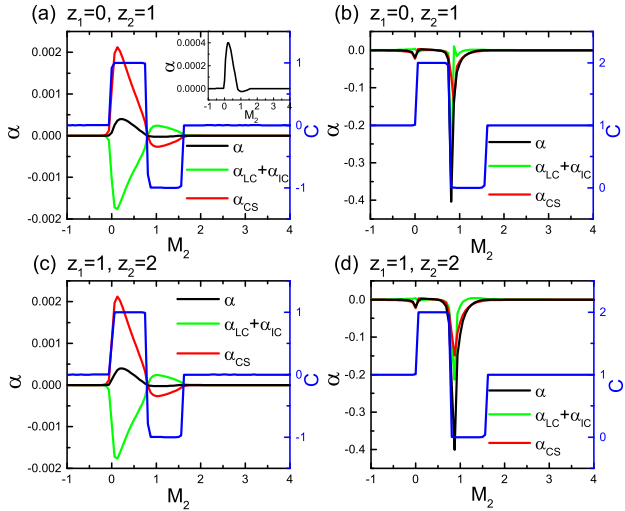}
	\caption{ The total OMP and its constituent terms with the whole sample average as a function of  $M_2$.
First row: coordinate $z_1=0$, $z_2=1$. Second row: $z_1=1$, $z_2=2$.
Left column: $M_1=-0.5$, and the Chern number $C=0+C_2$. Right column: $M_1=0.5$, and the Chern number $C=1+C_2$. Black line represents the total OMP $\alpha$. Red line represents the constituent term $\alpha_{\mathrm{CS}}$. Green line represents the constituent term $\alpha_{\mathrm{LC}}+\alpha_{\mathrm{IC}}$. Blue line represents the Chern number $C=C_1+C_2$ of the bilayer sample. The electric field $\varepsilon=0.001$, and other model parameters are the same as Fig. \ref{FigM-M2-Z1Z2}. }
	\label{Figalpha-z}	
\end{figure}

In Fig. \ref{Figalpha-z}, we show behaviors of the total OMP $\alpha$ and its constituent terms, $\alpha_{\mathrm{CS}}$ and $\alpha_{\mathrm{LC}}+\alpha_{\mathrm{IC}}$, by changing the parameter $M_2$, which directly controls the bulk gap of the second layer and therefore controls the Chern number $C_2$. For comparison, the first (second) row of panels correspond to the coordinate $z_1=0$ and $z_2=1$ ($z_1=1$, $z_2=2$), i.e., with a shift of $z$-coordinate.

For the case of Chern number $C=C_1+C_2=0+C_2$, by comparing Fig. \ref{Figalpha-z} (a) and (c), we can see that
the total OMP $\alpha$ and its constituent terms $\alpha_{\mathrm{LC}}+\alpha_{\mathrm{IC}}$, $\alpha_{\mathrm{CS}}$, remain unchanged under these such a shift of the $z$-coordinate.
This holds similarly for the case of $C=C_1+C_2=1+C_2$ as shown in Fig. \ref{Figalpha-z} (b) and (d). These results are consistent with previous section's conclusion that the OM and its constituents are gauge invariant, when they are averaged over the whole sample. Therefore correspondingly, the total OMP $\alpha$ and the constituent terms $\alpha_{\mathrm{LC}}+\alpha_{\mathrm{IC}}$, and $\alpha_{\mathrm{CS}}$, are also gauge invariant with the whole sample average.

Due to such a gauge invariance of OMP from the whole sample average, we can just fix the $z$-coordinate as $z_1=0$ and $z_2=1$ in the following. Let us first focus on the case of $C=0+C_2$ in Fig. \ref{Figalpha-z} (a).
The most prominent feature is the peaks of the magnitudes of $\alpha$, $\alpha_{\mathrm{LC}}+\alpha_{\mathrm{IC}}$ and $\alpha_{\mathrm{CS}}$ simultaneously around $M_2=0.05$, when the Chern number $C=0+C_2$ jumps from $0+0$ to $0+1$. Then, their magnitudes decreases gradually to zero at $M_2\sim 0.8$, where another topological phase transition towards $C=0-1$ happens. Subsequently, the signs of $\alpha$, $\alpha_{\mathrm{LC}}+\alpha_{\mathrm{IC}}$ and $\alpha_{\mathrm{CS}}$ are reversed respectively.
The magnitudes of $\alpha_{\mathrm{LC}}+\alpha_{\mathrm{IC}}$ and $\alpha_{\mathrm{CS}}$ reach another maximum and then decreases to zero at $M_2<1.55$. But this maximum value is much smaller than the peak at $M_2=0.05$.
If we further increase $M_2$, even though there is a topological phase transition of $C=0-1\rightarrow0+0$ at $M_2\sim 1.6$, $\alpha$, $\alpha_{\mathrm{LC}}+\alpha_{\mathrm{IC}}$ and $\alpha_{\mathrm{CS}}$ remain zero instead of showing any peak.

Our previous work find that a disorder induced topological transition tends to be accompanied by a remarkable enhancement of OMP\cite{Wang2019}. However, above results indicate that, although a topological phase transition usually causes a sudden change of OM, it is \emph{not} always accompanied by a change of OMP. This also applies to the case of $C=1+C_2$ in Fig. \ref{Figalpha-z} (b). As $M_2$ changes from $-1$ to $4$, the Chern number $C$ changes from $1+0$ to $1+1$, then to $1-1$, and finally back to $1+0$. $\alpha$, $\alpha_{\mathrm{LC}}+\alpha_{\mathrm{IC}}$ and $\alpha_{\mathrm{CS}}$ present a peak only at the first two topological phase transitions. It is worth noticing the magnitude of the $Y$ axis in Fig. \ref{Figalpha-z} (b) is nearly $100$ times larger than that in Fig. \ref{Figalpha-z} (a). Hence, both peaks of OMP for the $C=1+C_2$ case are larger than those for the $C=0+C_2$ case.

\begin{figure}[htbp]
	\includegraphics*[width=0.48\textwidth]{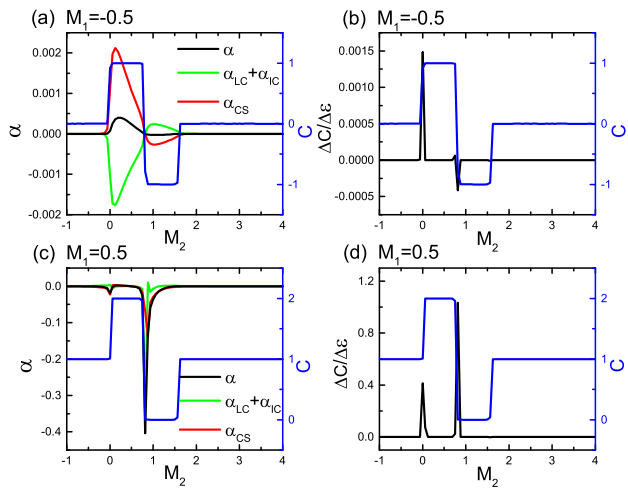}
	\caption{ Left column: copies of Fig. \ref{FigC-Efield} (a) and (b). Right column: the corresponding response of Berry curvature to electric field (black line) as a function of $M_2$. First row: $M_1=-0.5$, and the Chern number $C=0+C_2$. Second row: $M_1=0.5$, and the Chern number $C=1+C_2$. The electric field $\varepsilon=0.001$, and other model parameters are the same as Fig. \ref{FigM-M2-Z1Z2}. }
	\label{FigC-Efield}	
\end{figure}

Since a topological phase transition is not sufficient to produce a large value of OMP (CSOMP), it is necessary to clarify the origin and to identify those transitions that can greatly enhance the OMP (CSOMP). We will show that OMP (CSOMP) is related to the response of the integrated Berry curvature to the electric field. Let us define the response of Berry curvature under an external electric field as following,
\begin{equation}\label{EqdeltaC}
\frac{\triangle{C}}{\triangle{\varepsilon}}=\frac{C(\varepsilon)-C(0)}{\varepsilon}
\end{equation}
The resulting $\frac{\triangle{C}}{\triangle{\varepsilon}}$ corresponding to the cases of Fig. \ref{Figalpha-z} (a) and (b) [replotted as Fig. \ref{FigC-Efield} (a) and (c)] are presented as Fig. \ref{FigC-Efield} (b) and (d) respectively.
Fig. \ref{FigC-Efield} (b) shows the change of $\frac{\triangle{C}}{\triangle{\varepsilon}}$ (black line) and Chern number $C=0+C_2$ (blue line) vs $M_2$. By comparing Fig. \ref{FigC-Efield} (a) with (b), we can see that peaks of the total OMP $\alpha$ correspond to sharp peaks of $\frac{\triangle{C}}{\triangle{\varepsilon}}$. Besides, the magnitude of $\frac{\triangle{C}}{\triangle{\varepsilon}}$ with the topological transition $C=0+0\rightarrow0+1$
is larger than that with $C=0+1\rightarrow0-1$, as shown in Fig. \ref{FigC-Efield} (b).
Meanwhile, the peak value of $\alpha$ with the topological transition $C=0+0\rightarrow0+1$ is higher than that for $C=0+1\rightarrow0-1$ in Fig. \ref{FigC-Efield} (a).
Similar correspondences also appear in Fig. \ref{FigC-Efield} (b) and (d).
We have mentioned above that OMP in panel (b) is nearly $100$ times larger than in panel (a). Correspondingly the value of $\frac{\triangle{C}}{\triangle{\varepsilon}}$ in panel (d) is similarly $100$ times larger than that in panel (b).
In conclusion, there is a perfect correspondence (positions and magnitudes) between the peaks of OMP $\alpha$ and those of the differential Berry curvature
$\frac{\triangle{C}}{\triangle{\varepsilon}}$ at topological phase transition. This similarly applies to the CSOMP $\alpha_{\mathrm{CS}}$ term.

\begin{figure}[htbp]
	\includegraphics*[width=0.48\textwidth]{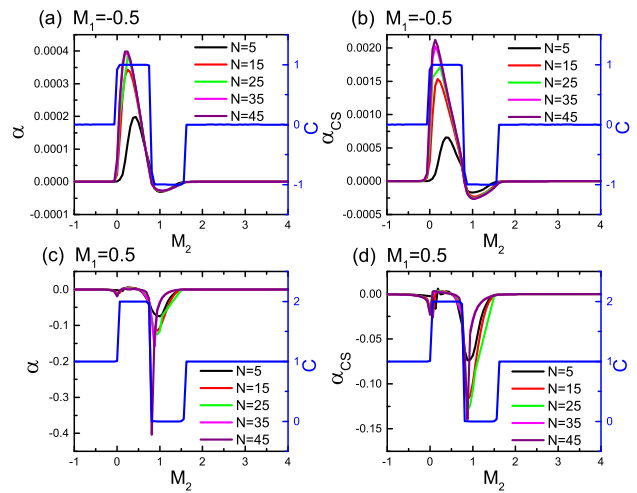}
	\caption{ Left column: the total OMP $\alpha$ with the whole sample average as a function of $M_2$ for different sample sizes $N$. Right column: the CSOMP term $\alpha_{\mathrm{CS}}$ as a function of $M_2$ for different sample sizes $N$. (a) and (b) $M_1=-0.5$, and the Chern number $C=0+C_2$. (c) and (d) $M_1=0.5$, and the Chern number $C=1+C_2$. The other model parameters are the same as Fig. \ref{FigM-M2-Z1Z2}. }
	\label{Figalpha-N}	
\end{figure}

The above $r$ space results are performed on finite size samples. Before ending, we discuss the size effects of our physical results. Fig. \ref{Figalpha-N} shows the total OMP $\alpha$ and the CSOMP term $\alpha_{\mathrm{CS}}$ with increasing sample sizes $N$ , for the case $C=0+C_2$ in panel (a) and (b), and the case $C=1+C_2$ in panel (c) and (d), respectively. In both cases, the peak shapes of both the OMP $\alpha$ and the CSOMP $\alpha_{\mathrm{CS}}$  become sharper with the increasing the sample sizes. These peaks seems to converge until the sample size $N\times{N}=45\times{45}$. This scaling growth confirms the stability of these peaks towards the thermodynamic limit, which further confirms these OMP and CSOMP peaks experimentally observable in realistic materials.

\section{V. Summary}
In summary, based on a bilayer model with an adjustable Chern number, we numerically investigated the orbital magnetization (OM) $M$ under a finite external electric field $\varepsilon$ with two different averages: averaging over the bulk or over the whole sample. Our first key finding is that the total OM $M$ over the bulk region has a gauge dependent on the the $z$ coordinate as $\triangle M_\mathrm{s}=\frac{\varepsilon}{2\pi}(C_1z_1+C_2z_2)$ where $C_1$ ($C_2$) represents the Chern number associated with the first (second) layer, and $z_1$ ($z_2$) is the corresponding $z$ coordinate. By scrutinizing the OM constituent terms, we find that this gauge shift comes from the gauge discontinuity of the Chern-Simons orbital magnetoelectric term $M_\mathrm{CS}$, which is  previously captured in $k$ space. Fortunately, this gauge shift of OM vanishes when the average is over the whole sample. This means that it is reliable to adopt the whole sample average of OM for a layered Chern insulator under nonzero electric field.

Based on the knowledge of OM with a finite electric field, we further investigate the OMP, in order to look for ways to significantly enhance OMP $\alpha$ and its constituent term, namely, the CSOMP $\alpha_\mathrm{CS}$.
The result shows that the OMP and CSOMP are generally enhanced at topological phase transitions, provided the response of Berry curvature to electric field is nonzero. The stronger the response of Berry curvature to electric field, the larger the OMP and CSOMP are.

\section{Acknowledgements}
This work was supported by National Natural Science
Foundation of China under Grant No. 12104108 and No. 11774336, the Joint Fund with Guangzhou Municipality under No. 202201020198, and the Starting Research Fund from
Guangzhou University under Grant No. RQ2020082 and No. 62104360.

\end{document}